\documentstyle[12pt,epsf]{article}

\epsfverbosetrue

\let\Huge=\huge
\let\huge=\Large
\let\Large=\large
\let\large=\normalsize

\newcommand{\cc}{c\bar{c}}
\newcommand{\bb}{b\bar{b}}
\newcommand{\QQ}{Q\bar{Q}}
\newcommand{\al}{\alpha}
\newcommand{\GeV}{{\rm GeV}}

\newcommand{\ms}{\overline{{\rm MS}}}

\newcommand{\be}{\begin{equation}}
\newcommand{\ee}{\end{equation}}
\newcommand{\bea}{\begin{eqnarray}}
\newcommand{\eea}{\end{eqnarray}}
\newcommand{\gtap}{{\raise.3ex\hbox{$>$\kern-.75em\lower1ex\hbox{$\sim$}}}}
\newcommand{\ltap}{{\raise.3ex\hbox{$<$\kern-.75em\lower1ex\hbox{$\sim$}}}}
\newcommand{\Dslash}{D\!\!\!\!\slash}
\newcommand{\bm}[1]{\mbox{\boldmath ${#1}$}}

\newcommand{\PRD}[3]{{\em Phys. Rev.} {\bf D{#1}} (19{#2}) {#3}}
\newcommand{\PRB}[3]{{\em Phys. Rev.} {\bf B{#1}} (19{#2}) {#3}}
\newcommand{\PRL}[3]{{\em Phys. Rev. Lett.} {\bf {#1}} (19{#2}) {#3}}
\newcommand{\PLB}[3]{{\em Phys. Lett.} {\bf B{#1}} (19{#2}) {#3}}
\newcommand{\NPB}[3]{{\em Nucl. Phys.} {\bf B{#1}} (19{#2}) {#3}}
\newcommand{\NPBproc}[3]{{\em Nucl. Phys.} {\bf B} (Proc. Suppl.)
           {\bf {#1}} (19{#2}) {#3}}
\newcommand{\ARNPS}[3]{{\em Annu. Rev. Nucl. Part. Sci.}
           {\bf #1} (19{#2}) {#3}}
\newcommand{\CMP}[3]{{\em Comm. Math. Phys.} {\bf {#1}} (19{#2}) {#3}}
\newcommand{\PRepC}[3]{{\em Phys. Rep.} {\bf C{#1}} (19{#2}) {#3}}

\def\Thebibliography#1{\section*{\centering REFERENCES}\list
{\arabic{enumi}.}{\settowidth\labelwidth{#1.}\leftmargin\labelwidth
 \advance\leftmargin\labelsep
 \usecounter{enumi}}}
 
 \sfcode`\.=1000\relax

\begin{document}
\renewcommand{\thefootnote}{\fnsymbol{footnote}}
\begin{flushright}
hep-ph/9509381 \\
September 1995 \\
\end{flushright}
\vspace{0.1cm}
\begin{center}
{\Huge\bf Charmonium Theory\footnote{Talk presented at the
Workshop on the Tau/Charm Factory, Argonne National Laboratory,
June 21--23, 1995.}}
\vspace{0.2in}
\end{center}
\begin{center}
{\Large Aida X. El-Khadra\footnote{Address after Sept. 1, 1995:
Physics Department, University of Illinois, 1110 W. Green St.,
Urbana, IL~61801}}
\end{center}
\vskip 0.2in
\begin{center}
{\small\em Physics Department, Ohio State University, 174 W 18th Ave,
Columbus, OH 43210, U.S.A.}
\vspace{0.3in}
\end{center}
\begin{quote}
\small{\bf Abstract.}
Recent theoretical progress in calculations of the spectrum and decays
of charmonium is reviewed.
Traditionally, our understanding of charmonium was based on
potential models. This is now being replaced by first principles.
\end{quote}
\vskip 1cm

\setcounter{footnote}{0}
\renewcommand{\thefootnote}{\alph{footnote}}

\section*{\centering INTRODUCTION} \label{sec:IM}

The charmonium system is
at present among the theoretically best understood hadronic systems.
The charm quark mass is large compared to the typical QCD
scale, $\Lambda_{QCD}$. The $\cc$ bound states are therefore
governed by non-relativistic dynamics. Historically, while the
QCD potential was not known from first principles,
relatively simple guesses for
phenomenological potentials had proven quite successful in describing
the experimentally measured bound state spectrum of charmonium \cite{quigg}.
Likewise, for annihilation decays of charmonium,
an intuitive factorization
{\em ansatz}, involving the wave function at the origin (calculable in
potential models) worked quite well, for the most part.

This model-based theoretical understanding is now being replaced
by first principles. For charmonium spectroscopy, the progress comes
from using lattice QCD. Charmonium decays can be treated rigorously
with the help of non-relativistic QCD (or NRQCD).

Lattice field theory offers a systematic first
principles approach to solving QCD.
It has been argued by Lepage \cite{lepage} that charmonium is one of
the easiest systems to study with lattice QCD, with the potential
of complete control over all systematic errors.
Finite-volume errors are much easier to control for
quarkonia than for light hadrons.
Lattice-spacing errors, on the other hand, can be larger
for quarkonia and need to be considered.
An alternative to reducing the lattice spacing in order to control
this systematic error is improving the action (and operators).
For quarkonia, the size of lattice-spacing errors in a numerical
simulation can be {\em anticipated} by calculating expectation
values of the corresponding operators using potential model wave
functions. They are therefore ideal systems to test and establish
improvement techniques.
Most of the work of phenomenological relevance is done in
what is generally referred to as the ``quenched''
(and sometimes as the ``valence'') approximation.
In this approximation gluons are not allowed to split into
quark - anti-quark pairs (sea quarks). In the case of
charmonium, potential model phenomenology can be used to
estimate this systematic error.
Control over systematic
errors in turn allows the extraction of Standard Model parameters
from the quarkonia spectra.

Non-relativistic systems, like charmonium, are best described
with an effective field theory, non-relativistic QCD (or NRQCD).
It was shown in Ref.~\cite{bbl} that the application of NRQCD
to the problem of annihilation decays of quarkonium leads to
a general factorization formula. It reproduces the earlier
({\em ad hoc}) factorization {\em ansatz} for S-wave decays while putting it
on a firm theoretical footing. In the case of P-wave decays, the
previous {\em ansatz} is modified.

Lattice QCD and NRQCD are introduced in the following subsections.
The remainder of the talk is organized in two parts.
The first part reviews recent progress in calculations of the
charmonium spectrum based on lattice QCD.
The second part reviews progress in understanding
charmonium decays based on NRQCD.

\subsection*{\centering An Introduction to Lattice QCD} \label{sec:primer}

Lattice Field theory is formulated using the Feynman path integral
in Euclidean space. The quantities that are actually calculated
are expectation values of Greens functions (${\cal G}$), which are
products of gauge and fermion fields. The physical quantities of interest,
hadron masses, matrix elements, etc., are then extracted from these
Greens functions.

The discretization of space-time (with lattice spacing $a$) regulates
the path integral at short distances or in the ultraviolet.
A finite volume (of length $L$) is necessary for numerical
techniques and also introduces an infrared cut-off or momentum-space
discretization.
The vacuum expectation of a Greens function, ${\cal G}$,
is defined as:
\be  \label{eq:lim}
\langle {\cal G} \rangle = \lim_{L \rightarrow \infty} \,
         \lim_{a \rightarrow 0} \, \langle {\cal G} \rangle_{L,a}
 \;\;,\;\;\;
\langle {\cal G} \rangle_{L,a} = Z^{-1}_{L,a}
\int {\cal D}\psi {\cal D}\bar{\psi} {\cal D}U \, {\cal G}
                                      \, e^{-S_{\rm lat}} \;\;\;.
\ee
$Z_{L,a}$ normalizes the expectation value.
I have omitted spin and color indices for compactness.
The gauge degrees of freedom are written as (path ordered) exponentials
of the gauge field, $A_{\mu}$:
\be
U_{\mu} (x) = e^{ i \int_x^{x+a} dx' A_{\mu} (x')} \simeq e^{ia A_{\mu} (x)}
    \;\;\;,
\ee
which makes it easy to maintain gauge invariance.
The link fields, $U$, are $SU(3)$ matrices.
The (Euclidean) QCD action,
\be       \label{eq:qcd}
S = S_g + S_f \;\;, \;\;\;\;
S_g = \frac{1}{4g^2} \int d^4x \, F_{\mu \nu}F^{\mu \nu} \;\;, \;\;\;\;
S_f = \int d^4x \, \bar{\psi}(x) (\Dslash + m ) \psi (x) \;\;.
\ee
is discretized, such that Eq.~(\ref{eq:qcd}) is recovered in the
the continuum ($a \rightarrow 0$) limit:
\be  \label{eq:limit}
 S_{\rm lat} = S + {\cal O} (a^n)
     \;\;, \;\;\; n \geq 1 \;\;\;.
\ee
I will not go into the explicit formulations of $S_{\rm lat}$
here, but instead refer the reader to pedagogical introductions~\cite{intro}.
The most common form for the gauge action is Wilson's \cite{wilson_g},
written in terms of plaquettes
-- products of $U$ fields around the smallest closed loop on a lattice.
Wilson's gauge action has discretization errors of ${\cal O}(a^2)$.

For fermions the situation is more complicated.
The discretization of
\be
M \equiv \Dslash + m \;\;,
\ee
is a sparse, finite dimensional matrix.
Two different approaches are in use.
In Wilson's formulation \cite{wilson_f} chiral symmetry is explicitly
broken, but restored in the continuum limit.
The pay-off is a solution of the so-called fermion doubling problem.
Staggered fermions \cite{ks} keep a $U(1)$ chiral symmetry
at the expense of dealing with 4 degenerate flavors of
fermions.

Eq.~(\ref{eq:lim}) emphasizes that QCD is a limit of lattice QCD.
However, in numerical calculations these limits cannot be
taken explicitly, only by extrapolation. This is feasible, because
theoretical guidance for both limits is available.
The zero-lattice-spacing limit is guided by asymptotic freedom, since
the lattice spacing is related to the gauge coupling by the
renormalization group.
Quantum field theories in large but finite volumes have also been
analyzed theoretically \cite{ml_vol}.

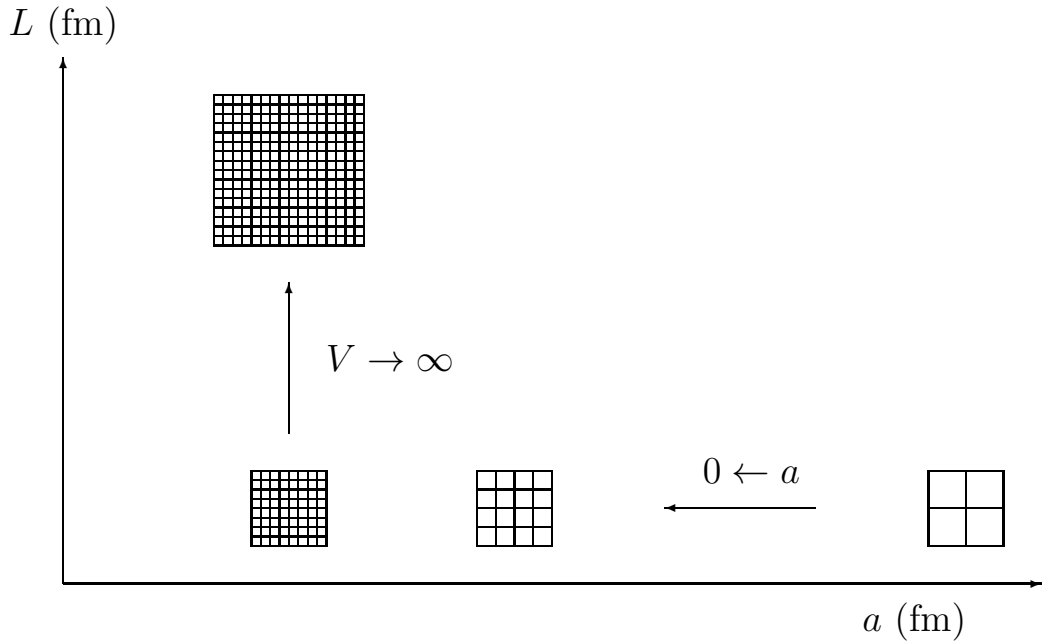
\begin{figure}[htb]
\begin{center}
\setlength{\unitlength}{1mm}
\begin{picture}(110,80)(0,-5)
\multiput(115,15)(5,0){3}{\line(0,-1){10}}
\multiput(115,15)(0,-5){3}{\line(1,0){10}}
\multiput(55,15)(2.5,0){5}{\line(0,-1){10}}
\multiput(55,15)(0,-2.5){5}{\line(1,0){10}}
\multiput(25,15)(1.25,0){9}{\line(0,-1){10}}
\multiput(25,15)(0,-1.25){9}{\line(1,0){10}}
\multiput(20,65)(1.25,0){17}{\line(0,-1){20}}
\multiput(20,65)(0,-1.25){17}{\line(1,0){20}}
\put(100,10){\vector(-1,0){20}}
\put(85,15){\makebox(0,0)[l]{\Large $0 \leftarrow a$}}
\put(30,20){\vector(0,1){20}}
\put(35,30){\makebox(0,0)[l]{\Large $V \rightarrow \infty$}}
\put(0,0){\vector(0,1){70}}
\put(0,0){\vector(1,0){130}}
\put(0,73){\makebox(0,0)[b]{\Large $L$ (fm)}}
\put(120,-5){\makebox(0,0)[r]{\Large $a$ (fm)}}
\end{picture}
\end{center}
\caption{Illustration of the continuum and infinite-volume limits.}
      \label{fig:lim}
\end{figure}

In a numerical calculation the limits are taken by considering
a series of lattices, as illustrated in Figure~\ref{fig:lim}.
While keeping the physical volume (or $L$) fixed, the lattice spacing is
successively reduced; then, keeping the lattice spacing fixed the
volume is increased. The calculation is in the continuum (infinite
volume) limit once the hadron spectrum or matrix elements of
interest become independent of the lattice spacing (volume).

In practice, however, limitations in computational resources
do not permit the ideal lattice QCD
calculation just described. In particular, the computational
cost of reducing
the lattice spacing naively scales like $(L/a)^4$. (The
computational cost is really higher, because of numerical problems
at smaller lattice spacings.)
Eq.~(\ref{eq:limit}) illustrates
an alternative. By improving the discretization errors in
the lattice action (and operators), the continuum limit
can be reached at coarser lattice spacings than before.
Simulations with improved actions can come at only a slightly
higher computational price.
The ideas underlying improvement were developed some time
ago \cite{sym,lw,sw}, and have since been
revitalized \cite{it,us_thy,perfect,adhlm}.

If the quark mass is large compared to the typical QCD scale,
$\Lambda_{QCD}$, effective theories (such as NRQCD) are most
adequate in describing the physics \cite{eff}. In that case, the lattice
spacing cannot be taken to zero. Lattice-spacing errors can, however, be
systematically reduced by improvement \cite{nrqcd_thy}.

The problem is now (more or less) set up. I again refer the reader
to the literature \cite{intro} for more details on the organization
of typical lattice QCD calculations.

\subsection*{\centering An Introduction to NRQCD (Non-Relativistic QCD)}

In non-relativistic systems, the velocity $v$ of the heavy quark
inside the bound state is a small parameter; in
charmonium $v^2 \sim 0.3$. The important momentum scales with
regard to the structure of the bound state are the heavy quark momentum
($mv$) and its kinetic energy ($mv^2$).
Momenta of the order of the heavy quark mass ($m$), or above, are
relatively unimportant in the bound state dynamics.
If $v$ is small enough, then all the different momentum
scales are well separated, in particular:
\be
\Lambda_{\rm QCD} \;, \; mv^2 \;,\; mv \ll m
\ee

The most adequate description of such systems is in terms of an
effective field theory, non-relativistic
QCD (or NRQCD). The theory has a cut-off $\Lambda \sim m$.
The effective lagrangian can be written as an expansion in powers of
$1/m$.
The NRQCD lagrangian is \cite{eff}
\be  \label{eq:eff}
{\cal L}_{\rm NRQCD} = {\cal L}_{\rm light} + {\cal L}_{\rm heavy}
   + \delta{\cal L} \;\;.
\ee
${\cal L}_{\rm light}$ is the fully relativistic lagrangian
for the light quarks and gluons.
${\cal L}_{\rm heavy}$ is the heavy quark (and anti-quark) lagrangian,
\be  \label{eq:nrqcd}
{\cal L}_{\rm heavy} = \psi^{\dagger}
           \left( iD_0 + \frac{\bm{D}}{2m} \right) \psi
         + \chi^{\dagger} \left( iD_0 - \frac{\bm{D}}{2m} \right) \chi
         \;\;,
\ee
where $\psi$ and $\chi$ are 2-component Pauli spinors describing quark and
anti-quark degrees of freedom.

Relativistic effects of full QCD introduce corrections to
Eq.~(\ref{eq:nrqcd}) which appear in $\delta{\cal L}$  as local,
non-renormalizable interactions, with coefficients that are
calculable in perturbation theory \cite{eff,nrqcd_thy}.
In principle, infinitely many terms must be considered to reproduce full
QCD.
In practice, however, only a finite number is needed,
since every operator scales with a certain
power of $v$, as shown in Ref.~\cite{nrqcd_thy}. These power
counting rules (e.g., $\psi \sim (mv)^{(3/2)}$, $\bm{D} \sim mv$, etc.)
effectively order the terms in the NRQCD lagrangian by powers of $v$.

The annihilation of a $\QQ$ pair occurs at momenta of order
$m$. This short distance physics cannot be treated directly in NRQCD.
However, the annihilation contribution to low-energy
$\QQ \rightarrow \QQ$ scattering can be incorporated in NRQCD by
adding local 4-fermion operators to $\delta {\cal L}$ \cite{bbl},
\be  \label{eq:4f}
\delta {\cal L}_{\rm 4-fermion} = \sum_i \frac{f_i}{m^{d_i -4}} \,
         {\cal O}_i \;\;.
\ee
For example,
\be
 {\cal O}_1 (^1S_0) = \psi^{\dagger} \chi \chi^{\dagger} \psi \;\;.
\ee
The $f_i$ are again calculable in perturbation theory as expansions
in $\al_s (m)$.

\section*{\centering CHARMONIUM SPECTROSCOPY} \label{sec:QQ}

Two different formulations for fermions have been used in lattice
calculations of these spectra.
Lepage and collaborators \cite{lepage,nrqcd_thy} have adapted the NRQCD
formalism to the lattice regulator. Several groups have performed
numerical calculations of quarkonia in this approach.
In Ref.~\cite{nrqcd_cc} the
NRQCD action is used to calculate
the charmonium spectrum, including
terms of ${\cal O} (mv^4)$ and ${\cal O}(a^2)$.
In addition, this group has calculated the $\bb$ spectrum in
the quenched approximation ($n_f=0$) \cite{nrqcd_spec} and also using gauge
configurations that include 2 flavors of sea quarks
\cite{nrqcd_als,nrqcd_mb}.

The Fermilab group \cite{us_thy} developed a generalization of previous
approaches, which encompasses the non-relativistic
limit for heavy quarks as well as Wilson's relativistic action
for light quarks. Lattice-spacing artifacts are analyzed for quarks with
arbitrary mass. Ref.~\cite{us} uses this approach to calculate
the $\cc$ (and $\bb$) spectra in the quenched approximation. We
considered the effect of reducing lattice-spacing errors from
${\cal O}(a)$ to ${\cal O}(a^2)$.

The two groups mentioned above use gauge configurations generated
with the Wilson action, leaving ${\cal O}(a^2)$ lattice-spacing errors
in the results. The lattice spacings, in this case, are in the range
$a \simeq 0.05 - 0.2$ fm.
Ref.~\cite{adhlm} uses an improved gauge action (to ${\cal O}(a^4)$)
together with a non-relativistic quark action improved to the same order
(but without spin-dependent terms) on coarse ($a \simeq 0.4 - 0.24$ fm)
lattices.

\begin{figure}[htb]
\begin{center}
\epsfxsize= 0.65\textwidth
\leavevmode
\epsfbox[11 219 588 560]{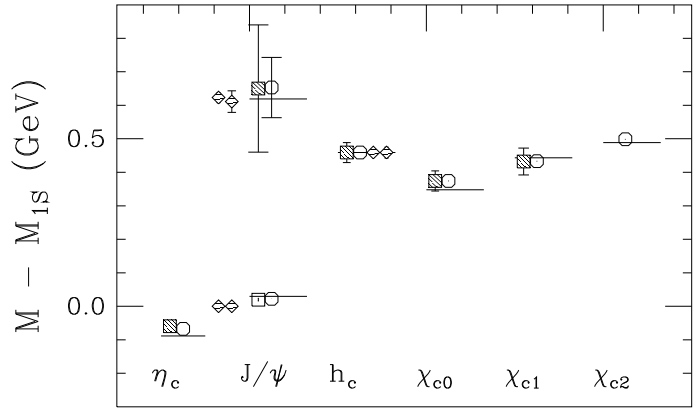}
\end{center}
\caption[xx]{A comparison of lattice QCD results for the $\cc$ spectrum
   using the quenched approximation ($n_f = 0$). The error bars
   are statistical only.
   ---: Experiment; $\Box$: FNAL \cite{us}; $\circ$: NRQCD \cite{nrqcd_cc};
   $\Diamond$: ADHLM \cite{adhlm}.}
 \label{fig:cc}
\end{figure}

The first step in any lattice QCD calculation is the determination of the
two free free parameters of the theory, the gauge coupling and quark
mass, from experiment. This is discussed in the following subsection.
The results for the charmonium spectrum from all groups are
summarized in Figure~\ref{fig:cc}.

The agreement between the experimentally-observed spectrum and
lattice QCD calculations is respectable. As indicated in the
preceding paragraphs, the lattice artifacts are different for all groups.
Figure~\ref{fig:cc} therefore emphasizes the level of
control over systematic errors.
I should also note, however, that the theoretical errors (from the
Monte Carlo integration alone) are still much larger than
the experimental ones. They also grow with every level of
excitation considered.

The first quarkonium results with 2 flavors of degenerate sea quarks
have appeared \cite{nrqcd_als,kek_als,cdhw} with lattice-spacing
and finite-volume errors similar to the quenched calculations,
significantly reducing this systematic error.
In Refs.~\cite{kek_als,cdhw} the 1P-1S splitting in
charmonium is calculated, while Ref.~\cite{nrqcd_als}
considers the $\bb$ spectrum.

Several systematic effects associated with the
inclusion of sea quarks must still be studied.
They include the dependence on the quarkonium
spectrum of the number of flavors of sea quarks and
the sea-quark action (staggered vs. Wilson). The inclusion
of sea quarks with realistic light-quark masses is very
difficult. However, quarkonia are expected to depend only
very mildly on the masses of the light quarks. This systematic
error has not been included yet and should be checked
numerically.

The first (and second) generation of lattice QCD calculations,
as described here,
is focused on the simplest physical quantities, like the
low lying states or simple (decay) matrix elements, to establish
the method. Once first principles calculations have been achieved,
this technology can and will (given sufficient motivation) be
used to look at more complicated problems,
like higher excited states, mixing with glueballs, hybrids, etc.

\subsection*{\centering Standard Model Parameters from
    Charmonium} \label{sec:sm}

The first step, the determination of the lattice gauge coupling and
quark mass, follows from comparing appropriate quantities calculated
on the lattice with the corresponding experimental measurements.
The lattice parameters can then be converted to their counterparts
in continuum QCD with perturbation theory.

Precise determinations of Standard Model parameters
are an interesting by-product of lattice QCD calculations of
charmonium (and bottomonium). However, the theoretical uncertainties
must be reduced by an order of magnitude before they become comparable
to the present experimental errors.
After discussing the determination
of the lattice spacing (which sets the scale in Figure~\ref{fig:cc}),
I will summarize the results for the strong coupling and
the charm quark mass from charmonium.

\subsubsection*{\it\centering Determination of the Lattice Spacing, $a$}

\begin{figure}[htb]
\begin{center}
\epsfxsize= 0.65\textwidth
\leavevmode
\epsfbox{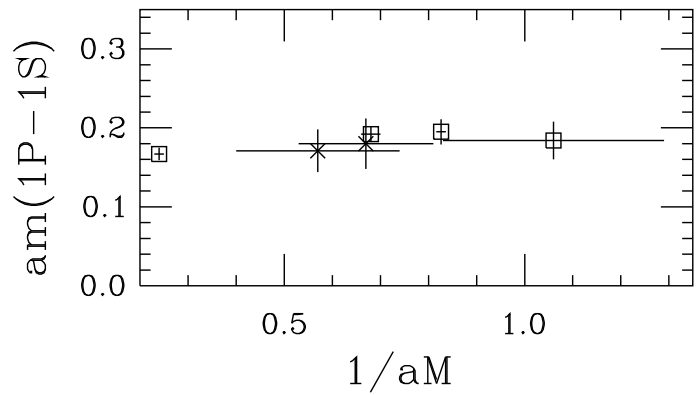}
\end{center}
\caption[xx]{The 1P-1S splitting as a function of the 1S mass
   (statistical errors only) from Ref. \cite{us};
   $\Box$: ${\cal O}(a^2)$ errors; $\times$: ${\cal O}(a)$ errors.}
 \label{fig:1p1s}
\end{figure}

\begin{figure}[htb]
\begin{center}
\epsfxsize= 0.65\textwidth
\leavevmode
\epsfbox{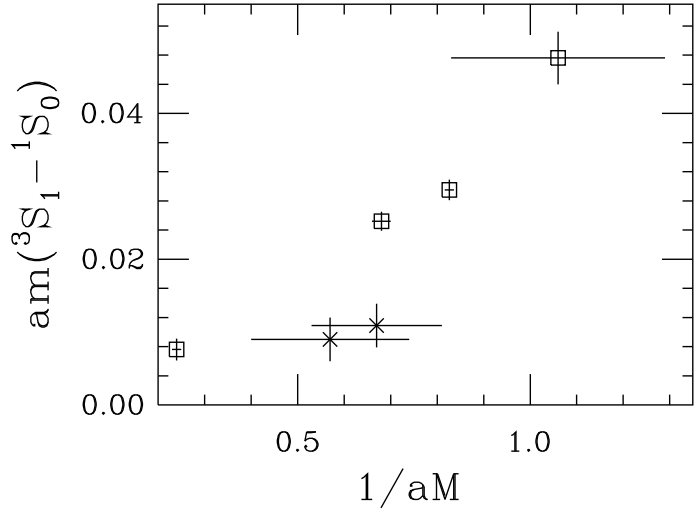}
\end{center}
\caption[xx]{The hyperfine splitting as a function of the 1S mass
   (statistical errors only) from Ref. \cite{us};
   $\Box$: ${\cal O}(a^2)$ errors; $\times$: ${\cal O}(a)$ errors.}
 \label{fig:hyp}
\end{figure}

The input gauge coupling sets the lattice spacing, $a$, which is
determined in physical units by comparing a suitable quantity on
the lattice with its experimental value.
For this purpose, one should identify quantities that are insensitive
to lattice errors. In quarkonia, spin-averaged splittings are good
candidates. The experimentally observed 1P-1S and 2S-1S splittings
depend only mildly on the quark mass (for masses between $m_b$ and $m_c$),
as shown in table~\ref{tab:QQexp}.
\begin{table}
\caption{Spin-averaged splittings in the $J/\psi$ and $\Upsilon$
systems in comparison.} \label{tab:QQexp}
\begin{center}
\begin{tabular}{rrrr}
\hline
  & $\cc$ (MeV) & $\bb$ (MeV) \\ \hline
$m({\rm 1P-1S})$  & $456.8$  & $452\;\;\;$ \\
$m({\rm 2S-1S})$  & $596\;\;\;$ & $563\;\;\;$ \\
$m({\rm 2P-1P})$  & ---$\;\,$  & $359.7$ \\ \hline
\end{tabular}
\end{center}
\end{table}
Figure~\ref{fig:1p1s}
shows the observed mass dependence of the 1P-1S splitting
in a lattice QCD calculation. The comparison between results from
different lattice actions illustrates that
higher-order lattice-spacing errors for these splittings
are small\cite{nrqcd_als,us}.
In contrast, Figure~\ref{fig:hyp} shows the hyperfine splitting
as an example of a quantity that strongly depends on both the mass
and the lattice action. It would therefore be a poor choice for a
determination of the lattice spacing.

\subsubsection*{\it\centering The Strong Coupling, $\al_s$}

Within the framework of lattice QCD the conversion from the bare
to a renormalized coupling can, in principle, be made
non-perturbatively \cite{luesch}.
An alternative is to define a renormalized coupling through short
distance lattice quantities \cite{lm}.
The size of higher-order corrections associated with the above
defined coupling constant can be tested by comparing
perturbative predictions for short-distance lattice quantities
with non-perturbative results \cite{lm}.

At this point the relation to the $\ms$ coupling is known
to 1-loop, leading to a $5 \,\%$ uncertainty. It has recently
been calculated to 2-loops \cite{lw-2l}
in the quenched  approximation (no sea quarks, $n_f = 0$).
The extension to $n_f \neq 0$ will significantly reduce
the uncertainty due to the use of perturbation theory.

{\em Sea Quark Effects.\  }
Calculations that properly include all sea-quark effects
do not yet exist.
If we want to make contact with the ``real world'', these effects
have to be estimated phenomenologically or extrapolated away.

The phenomenological correction necessary to account for
the sea-quark effects omitted in calculations of quarkonia
that use the quenched approximation gives rise to the dominant
systematic error in these calculations \cite{prl,nrqcd_l93}.
Similar ideas were used to
correct for sea-quark effects in early calculations of quarkonia
spectra from the heavy-quark potential calculated in quenched
lattice QCD \cite{rebbi}.

By demanding that, say, the spin-averaged 1P-1S splitting calculated on
the lattice reproduce the experimentally observed one (which
sets the lattice spacing, $a^{-1}$, in physical units), the effective
coupling of the quenched potential is in effect matched to the
coupling of the effective 3 flavor potential at the typical
momentum scale of the quarkonium states in question. The difference
in the evolution of the zero flavor and 3,4 flavor couplings
from the effective low-energy scale to the ultraviolet cut-off,
where $\al_s$ is determined, is the perturbative estimate
of the correction.

For comparison with other determinations of $\al_s$, the $\ms$
coupling can be evolved to the $Z$ mass scale. An average \cite{als_rev} of
Refs.~\cite{us,nrqcd_l93} yields for $\al_s$ from calculations
in the quenched approximation:
\be          \label{eq:nf0}
   \al^{(5)}_{\ms} (m_Z) = 0.110 \pm 0.006 \;\;\;.
\ee
The experimental error in this determination from the quarkonium
mass splitting is much smaller than the theoretical uncertainty
and does not contribute to the total.

The phenomenological correction described in the previous paragraphs
has been tested from first principles in
Refs.~\cite{nrqcd_als,kek_als,cdhw}. All groups calculate quarkonium
splittings with 0 and 2 flavors of sea quarks. After extrapolating
to the physical 3 flavor case and evolving the coupling to
$m_Z$, Refs.~\cite{kek_als,cdhw} find for the strong coupling from
charmonium
\be
 \al^{(5)}_{\ms} (m_Z) = 0.111 \pm 0.005
\ee
in good agreement with the previous result in Eq.~(\ref{eq:nf0}).
The total error is now dominated by the rather large
statistical errors and the perturbative uncertainty.

At present, the result of Ref.~\cite{nrqcd_als} has the
smallest statistical and systematic errors for the strong
coupling (in this case from the $\bb$ spectrum):
\be        \label{eq:ms5}
\al^{(5)}_{\ms} (m_Z) = 0.115 \pm 0.002 \;\;\;.
\ee

Phenomenological corrections are a necessary evil that enter most
coupling constant determinations.
In contrast, lattice QCD calculations with complete control over
systematic errors will yield truly first-principles determinations
of $\al_s$ from the experimentally observed hadron spectrum.

At present, determinations of $\al_s$ from the experimentally measured
quarkonia spectra using lattice
QCD are comparable in reliability and accuracy to other determinations
based on perturbative QCD from high energy experiments. They
are therefore part of the 1994 world average for $\al_s$ \cite{als_rev}.
The phenomenological corrections for the most important sources
of systematic errors in lattice QCD calculations of quarkonia are
now being replaced by first principles, which will significantly increase
the accuracy of $\al_s$ determinations from quarkonia.
In particular, the systematic errors associated with the
inclusion of sea quarks into the simulation have to be checked.

\subsubsection*{\it\centering The Charm Quark Mass} \label{sec:mass}

Because of confinement, the quark masses cannot be measured
directly, but have to be inferred from experimental
measurements of hadron masses, and depend on the calculational
scheme employed.
In lattice QCD quark masses are determined non-perturbatively,
by tuning the input lattice quark mass ($m^{\rm lat}_Q$) so that,
for example, the experimentally observed $J/\psi$ mass is reproduced
by the calculation.

Phenomenologically useful quark masses are the perturbatively defined
pole and $\ms$ masses, which the bare lattice mass can be related
to by perturbation theory:
\be  \label{eq:mms}
  m_Q^{\rm pole} = Z^{\rm pole}_m \, m^{\rm lat}_Q \;\;\;,\hspace{2cm}
  m_Q^{\ms} (m_Q) = Z^{\ms}_m \, m^{\rm lat}_Q \;\;\;.
\ee
Of course, as always, all systematic errors arising from the
lattice QCD calculation need to be under control for a
phenomenologically interesting result;
in particular, the systematic error introduced by the
(partial) omission of sea quarks has to be removed.
The short-distance corrections that introduced the dominant
uncertainty to the $\alpha_s$ determination from quarkonia
are absent for the pole mass determination, because this
effective mass does not run for momenta below its mass.
An analysis of the $b$-quark mass from the $\bb$ spectrum
with and without sea quarks is consistent with this
estimate \cite{nrqcd_mb}.

Ref.~\cite{us_mc} analyzes the the charm quark mass
from the charmonium spectrum with the preliminary result,
$m_c^{\rm pole} = 1.5 (2)$ GeV.

The $\ms$ mass for the charm quark has also been determined
from a compilation of $D$ meson calculations in the quenched
approximation \cite{elc_mq}, with
$m_c^{\ms} (2 \,\GeV) = 1.47 (28)$ GeV. The error includes
statistical errors from the original calculations and the
perturbative error. However sea-quark effects cannot, in this case,
be estimated phenomenologically, leaving this systematic
error uncontrolled.

\section*{\centering CHARMONIUM DECAYS} \label{sec:decays}

Historically \cite{itep,quigg}, charmonium annihilation decays were treated
with a factorization {\em ansatz}, which divides the decay rate into
a short distance, perturbative part, and a long distance,
non-perturbative part, usually parametrized as the wave
function at the origin, $R(0)$.
This {\em ansatz} worked quite well for S-wave decays,
even after including radiative corrections at next-to-leading order
\cite{s-decays}.
For P-wave decays, the long distance piece was
identified as the derivative of the wave function at the origin,
$R'(0)$.
However, the radiative corrections were found to have infra-red
divergences \cite{p-decays} for $J=1$ P-wave states already at leading order,
and for $J=0,2$ P-wave states at next-to-leading order,
preventing reliable theoretical predictions.
This {\em ansatz} also did not allow for a systematic inclusion of higher
Fock states, or relativistic corrections.

If the effective field theory framework of NRQCD is used to describe
annihilation decays, it was shown \cite{bbl} that a general factorization
formula holds for the decay rates.
The annihilation decay rate of a quarkonium state $A$ can
be written as
\be
 \Gamma ( A ) = 2 \; {\rm Im} \langle A | \delta {\cal L}_{\rm 4-fermion}
                 | A \rangle \;\;.
\ee
Using Eq.~(\ref{eq:4f}) this gives
\be   \label{eq:bbl}
\Gamma ( A ) = \sum_{i} \frac{F_i}{m^{d_i -4}}
             \; \langle A | {\cal O}_i | A \rangle \;\;\;.
\ee
The coefficients $F_i$ are the imaginary parts of the $f_i$ in
Eq.~(\ref{eq:4f}). Because of the power counting rules
of Ref.~\cite{nrqcd_thy}, the matrix elements scale with some power
of the velocity,
\be
\langle A | {\cal O}_i | A \rangle  \sim v^{2n} \;\;,
\ee
which usually increases as higher dimensional
operators are considered in the sum of Eq.~(\ref{eq:bbl}).
Effectively, this approach expresses the quarkonium decay
rates as expansions in the the short distance parameter $\al_s(m)$
and the long distance parameter $v^2$.
The desired accuracy of the theoretical prediction thus serves
as a truncation criterion for the (infinite) sum in Eq.~(\ref{eq:bbl}).

The matrix elements in Eq.~(\ref{eq:bbl}) are calculable from first
principles using lattice NRQCD. These calculations are in progress
by the ANL group \cite{anl}. In the meantime,
the matrix elements can also be extracted from experimental measurements
of charmonium decay rates.

It was shown in Ref.~\cite{bbl} that heavy-quark spin symmetry and
the vacuum-saturation approximation reduces the number of independent
matrix elements that have to be determined non-perturbatively or
phenomenologically.

The matrix elements of the 4-fermion operators,
which contribute to the  charmonium decays into light hadrons
can be simplified using the vacuum saturation approximation.
This approximation is valid in NRQCD up to relative order $v^4$.
For example,
\be  \label{eq:vac}
 \langle \eta_c | \psi^{\dagger} \chi \chi^{\dagger} \psi | \eta_c \rangle =
 |\langle 0 | \chi^{\dagger} \psi | \eta_c \rangle |^2
 ( 1 + {\cal O}(v^4) ) \;\;.
\ee
This relates the matrix elements appearing in electro-magnetic
charmonium decays to those in strong decays (into light hadrons).

Heavy-quark spin symmetry relates matrix elements of states in
the same radial and orbital levels but different spins. For example,
\be  \label{eq:spin}
  \bm{\epsilon}^* \cdot
\langle 0 | \chi^{\dagger} \bm{\sigma} \psi | J/\psi \rangle
  =  \langle 0 | \chi^{\dagger} \psi | \eta_c \rangle
  ( 1 + {\cal O}(v^2) ) \;\;.
\ee

The following two subsections discuss the application
of this approach to S- and P-wave decays, using specific
examples.

\subsection*{\centering S-Wave Decays}

I discuss the theoretical knowledge of S-wave decays using
$\eta_c \rightarrow \gamma\gamma$ as an example.
At leading order in $v^2$ the rate is
\be  \label{eq:s-v2}
\Gamma ( \eta_c \rightarrow \gamma\gamma )
  = \frac{F_{\gamma\gamma} (^1S_0)}{m^2} |\overline{R_S}|^2 \;\;,
\ee
where by spin symmetry (see Eq.~(\ref{eq:spin})) $\overline{R_S}$
can be taken as
$\overline{R_{\eta_c}}$ or the spin-averaged combination of
$\overline{R_{\eta_c}}$ and $\overline{R_{\psi}}$, with
\be
 \overline{R_{\eta_c}} = \sqrt{\frac{2\pi}{3}} \;
   \langle 0 | \chi^{\dagger} \psi | \eta_c \rangle \;\;.
\ee
The other $\eta_c$ and $J/\psi$ decays differ from Eq.~(\ref{eq:s-v2})
in the short distance coefficient but, using Eqs.~(\ref{eq:vac})
and (\ref{eq:spin}),
have the same long distance parameter, $\overline{R_S}$.
Identifying $\overline{R_S}$ with the wave function at the origin,
$R_S(0)$, leads back to the old factorization {\em ansatz}.
The short distance coefficients, $F_i$, are known to next-to-leading
order in $\al_s$.

At next to leading order in $v^2$, the decay rate becomes
\be
\Gamma ( \eta_c \rightarrow \gamma\gamma )
  = \frac{F_{\gamma\gamma} (^1S_0)}{m^2} \; |\overline{R_{\eta_c}}|^2
  + \frac{G_{\gamma\gamma} (^1S_0)}{m^4} \;
  {\rm Re}(\overline{R_S}^* \overline{\bm{\nabla}^2R_S}) \;\;.
\ee
Now the difference between $\overline{R_{\eta_c}}$ and $\overline{R_S}$,
which is of order $v^2$ (see Eq.~(\ref{eq:spin})) has to be taken into
account for consistency.
$\overline{\bm{\nabla}^2R_S}$ can be interpreted as the laplacian
of the wave function at the origin,
\be
 \langle 0 | \chi^{\dagger} (\bm{D}^2) \psi | \eta_c \rangle =
 \sqrt{\frac{3}{2\pi}} \;  \overline{\bm{\nabla}^2R_S}
 \;  ( 1 + {\cal O}(v^2) )  \;\;.
\ee
The short distance coefficients, $G_i$, are known to leading
order in $\al_s$.

The long distance parameters, $\overline{R_{\eta_c}}$,
$\overline{R_{\psi}}$ and $\overline{\bm{\nabla}^2R_S}$ can be
calculated in lattice NRQCD, estimated using potential models
or extracted from a phenomenological analysis of experimental data.
The relative errors (from the truncation of the perturbative and
non-relativistic series) are $\al_s(m_c)^2$, $\al_s (m_c) v^2$
and $v^4$, adding up to $\ltap 15 \;\%$.

It is not inconceivable that at least for the electro-magnetic
S-wave decays some (or all) of the higher order calculations
will be performed, reducing the theoretical error accordingly.

\subsection*{\centering P-Wave Decays}

The structure of P-wave decay rates is more complicated than
that of S-waves. I shall start with a simple example first,
the decay $\chi_{cJ} \rightarrow \gamma \gamma$.
\be
\Gamma ( \chi_{cJ} \rightarrow \gamma \gamma ) =
  \frac{ F_{\gamma \gamma}( ^3P_J)}{m^4} \; |\overline{R'_P}|^2 \;\;,
\ee
where again, by spin symmetry $\overline{R'_P}$ can be taken
from the matrix elements of any of the P-wave states (or their
spin-averaged combination). For example,
\be
   \langle 0 | \chi^{\dagger} (\frac{1}{2}\bm{D\cdot \sigma}) \psi
             | \chi_{c0} \rangle =
  \sqrt{\frac{27}{2\pi}} \;  \overline{R'_{\chi_{c0}}} \;
 (1 + {\cal O} (v^2) ) \;\;.
\ee
The obvious interpretation of $\overline{R'_P}$ as the derivative
of the wave function at the origin, $R'_P(0)$, again connects this
formalism to the old factorization {\em ansatz}.
The coefficients, $F_{\gamma\gamma}$, are known to next-to-leading
order in $\al_s$. The relative error is $v^2$ or $\sim 25\, \%$ of the
decay rate.

The situation is different for strong decays of
P-wave states. Taking as an example $h_c \rightarrow LH$ (light
hadrons), Eq.~(\ref{eq:bbl}) gives at leading order in $v^2$,
\be
\Gamma ( h_c \rightarrow LH)
  = \frac{F_1 (^1P_1)}{m^4}   \langle {\cal O}_1 \rangle
  + \frac{F_8 (^1S_0)}{m^4}  \langle {\cal O}_8 \rangle  \;\;,
\ee
where
\bea
  \langle {\cal O}_1 \rangle & = &
   \langle h_c | \psi^{\dagger} (\frac{1}{2} \bm{D \cdot \sigma}) \chi
       \chi^{\dagger} (\frac{1}{2} \bm{D \cdot \sigma}) \psi | h_c
       \rangle  \\
& = & \frac{9}{2 \pi} \; |\overline{R'_P}|^2 \; ( 1 + {\cal O}(v^2) )
\nonumber
\eea
and
\be
  \langle {\cal O}_8 \rangle =
   \langle h_c | \psi^{\dagger} T^a \chi \chi^{\dagger}T^a \psi | h_c \rangle
   \;\;.
\ee
The two matrix elements $\langle {\cal O}_1 \rangle$ and
$\langle {\cal O}_8 \rangle$ enter at the same order in $v^2$, because
they are both suppressed by $v^2$ with respect to the leading order
S-wave matrix elements. ${\cal O}_1$ is a dimension eight operator; the two
powers of $\bm{D}$ give the $v^2$ suppression.
${\cal O}_8$ is of dimension
six, but its matrix element picks the
$|\cc g\rangle$ Fock state, where the $\cc$ pair is in a color-octet $^1S_0$
state. The dominant Fock state of charmonium is the color-singlet
$|\cc \rangle$.
\be
| A \rangle = \Psi_{\cc} | \cc \rangle + \Psi_{\cc g} | \cc g \rangle +
\ldots \;\;,
\ee
with $\Psi_{\cc g} \sim {\cal O}(v)$. This gives the $v^2$ suppression.

This departure from the old factorization {\em ansatz} solves the problem
of infrared divergences, and thus leads to a consistent treatment of
this case similar to the other charmonium decays.

The short distance coefficients, $F_1$, are known to next-to-leading
order, while the $F_8$'s are only known to leading order in $\al_s$.
At present, the dominant relative error is thus $\al_s$ and $v^2$,
adding up to $\sim 40 \, \%$.

It should be possible to
extend the (perturbative) calculations of P-wave decays in this
frame work to the same order as what is presently available
for S-wave decays, reducing the theoretical error to $\ltap 15 \, \%$
(assuming knowledge of the relevant matrix elements).

A comparison of theory and experiment for P-wave decays
shows fair agreement, albeit with rather large errors
from both sides \cite{mangano}.

\section*{\centering CONCLUSIONS} \label{sec:con}

Quarkonia were, upon their discovery, called the hydrogen
atoms of particle physics. Their non-relativistic nature
justified the use of potential models, which gave a nice,
phenomenological understanding of these systems.
This phenomenology is at present useful to control
systematic errors in lattice QCD calculations of the charmonium
spectrum. However, we are quickly moving towards
truly first-principles calculations of quarkonia using
lattice QCD, thereby testing QCD non-perturbatively.
In this sense, quarkonia are still the hydrogen atoms of
particle physics.
Precise determinations of the Standard Model parameters,
$\alpha_s$, $m_c$ (and $m_b$), are by-products of this work.

Still lacking for a first-principles result is the
proper inclusion of sea quarks. The most difficult
problem in this context is the inclusion of sea quarks
with physical light quark masses. At present, this can
only be achieved by extrapolation (from $m_q \simeq 0.3 - 0.5 m_s$
to $m_{u,d}$).
If the light quark mass dependence of the quarkonia spectra
is mild, as anticipated, the associated systematic error
can be controlled.
First-principles calculations of quarkonia could then be
performed with currently available computational resources.

The present theoretical status of charmonium annihilation
decays is rather promising. The frame-work developed in Ref.~\cite{bbl}
leads to a systematic expansion in $\al_s(m_c)$ and $v^2$, with
controllable uncertainties. Until first-principles calculations
of the non-perturbative matrix elements become available, this
formalism can still be tested phenomenologically, using experimental
data. Theoretical predictions for
the decay rates should be available with uncertainties of
$ \ltap 10 \, \%$, in most cases before the Tau/Charm factory
turns on.

It is conceivable, that by the time a Tau/Charm factory turns
on, the theory of charmonium will be solidly based upon first
principles with accurate predictions for the spectrum and decays
of the low-lying states.
We will have moved to the next stage in theoretical (first-principles)
calculations concerning, for example, the properties of hybrid states
or mixing with glueballs.
Experimental information gathered at the Tau/Charm
factory and earlier experiments \cite{ginsburg} will then
give us {\em precision} tests of perturbative and non-perturbative QCD
in the charmonium system.

\section*{\centering ACKNOWLEDGEMENTS}

I thank the organizers for an enjoyable conference, and
G. Bodwin, E. Braaten, A. Kronfeld, P. Lepage, P. Mackenzie,
C. Quigg and J. Shigemitsu for discussions while preparing
this talk.

\begin{Thebibliography}{9}
\bibitem{quigg} W. Kwong, J. Rosner and C. Quigg, \ARNPS{37}{87}{325}.
\bibitem{lepage} P. Lepage, \NPBproc{26}{92}{45};
       B. Thacker and P. Lepage, \PRD{43}{91}{196};
       P. Lepage and B. Thacker, \NPBproc{4}{88}{199}.

\bibitem{bbl} E. Braaten, G. Bodwin and P. Lepage, \PRD{46}{92}{1914};
       \PRD{51}{95}{1125}; E. Braaten, NUHEP-TH-94-22, hep-ph/9409286.

\bibitem{intro} For pedagogical introductions to Lattice Field Theory,
       see, for example:
       M. Creutz, {\em Quarks, Gluons and Lattices} (Cambridge
       University Press, New York 1985);
       A. Hasenfratz and P. Hasenfratz, \ARNPS{35}{85}{559};
       A. Kronfeld, in {\em Perspectives in the Standard Model},
       R. Ellis, C. Hill and J. Lykken (eds.) (World Scientific,
       Singapore 1992), p. 421;
       see also A. Kronfeld and P. Mackenzie, \ARNPS{43}{93}{793};
       A.~El-Khadra, in {\em Physics in Collision 14}, S. Keller and
       H. Wahl (eds.) (Editions Frontieres, Cedex - France 1995), p. 209;
       for introductory reviews of lattice QCD.
\bibitem{wilson_g} K. Wilson, \PRD{10}{74}{2445}.
\bibitem{wilson_f} K. Wilson, in {\em New Phenomena in Subnuclear Physics},
       A. Zichichi (ed.) (Plenum, New York 1977).
\bibitem{ks} L. Susskind, \PRD{16}{77}{3031};
       T. Banks, J. Kogut and L. Susskind, \PRD{13}{76}{1043}.
\bibitem{ml_vol} M. L\"{u}scher, \CMP{104}{86}{177};
       \CMP{105}{86}{153}.
\bibitem{sym} See for example, T. Bell and K. Wilson, \PRB {11}{75}{3431};
       K. Symanzik, \NPB{226}{83}{187}; {\em ibid.} 205.
\bibitem{lw} P. Weisz, \NPB{212}{83}{1}; M. L\"{u}scher and P. Weisz,
       \NPB {212}{84}{349}; \CMP{97}{85}{59}; (E) {\bf 98} (1985) 433.
\bibitem{sw} B. Sheikholeslami and R. Wohlert, \NPB{259}{85}{572}.
\bibitem{it} C. Heatlie, {\em et al.}, \NPB {352}{91}{266}.
\bibitem{us_thy} P. Mackenzie, \NPBproc{30}{93}{35};
       A. Kronfeld, \NPBproc{30}{93}{445};
       A. El-Khadra, A. Kronfeld and P. Mackenzie,
       Fermilab PUB-93/195-T.
\bibitem{perfect} P. Hasenfratz, \NPBproc {34}{94}{3};
       P. Hasenfratz and F. Niedermayer, \NPB{414}{94}{785};
       U. Wiese, \PLB{315}{93}{417};
       W. Bietenholz and U. Wiese, \NPBproc{34}{94}{516}.
\bibitem{adhlm} P. Lepage, in {\em The Building Blocks of Creation},
       S. Raby and T. Walker (eds) (World Scientific, Singapore 1994),
       hep-lat/9403018;
       M. Alford, {\em et al.}, \NPBproc{42}{95}{787}; hep-lat/9507010.
\bibitem{eff} E. Eichten and F. Feinberg, \PRD{23}{81}{2724};
       W. Caswell and P. Lepage, \PLB{167}{86}{437}.
\bibitem{nrqcd_thy} P. Lepage, {\em et al.}, \PRD{46}{92}{4052}.

\bibitem{nrqcd_cc} C. Davies, {\em et al.}, hep-lat/9506026.
\bibitem{nrqcd_spec} C. Davies, {\em et al.}, \PRD{50}{94}{6963}.
\bibitem{nrqcd_als} C. Davies, {\em et al.}, \PLB{345}{95}{42}.
\bibitem{nrqcd_mb} C. Davies, {\em et al.}, \PRL{73}{94}{2654}.
\bibitem{us} A. El-Khadra, G. Hockney, A. Kronfeld, P. Mackenzie,
       T. Onogi and J. Simone, Fermilab PUB-94/091-T.
\bibitem{kek_als} S. Aoki, {\em et al.}, \PRL{74}{95}{22}.
\bibitem{cdhw} M. Wingate, {\em et al.}, hep-lat/9501034.
\bibitem{luesch} For a review of $\alpha_s$ from the heavy-quark
       potential, see K. Schilling and G. Bali, \NPBproc{34}{94}{147};
       M. L\"{u}scher, R. Sommer, P. Weisz, and U. Wolff,
       \NPB{413}{94}{481}; G. de Divitiis, {\em et al.}, \NPB{433}{95}{390};
       \NPB{437}{95}{447};
       C. Bernard, C. Parrinello and A. Soni, \PRD{49}{94}{1585}.
\bibitem{lm} P. Lepage and P. Mackenzie, \PRD{48}{92}{2250}.
\bibitem{lw-2l} M. L\"{u}scher and P. Weisz, \PLB{349}{95}{165};
       hep-lat/9505011.
\bibitem{prl} A. El-Khadra, G. Hockney, A. Kronfeld and P. Mackenzie,
       \PRL{69}{92}{729}; A. El-Khadra, \NPBproc{34}{94}{141}
\bibitem{nrqcd_l93} The NRQCD Collaboration, \NPBproc{34}{94}{417}.
\bibitem{rebbi} D. Barkai, K. Moriarty and C. Rebbi, \PRD{30}{84}{2201};
       M. Campostrini, \PLB{147}{84}{343}.
\bibitem{als_rev} For reviews on the status of $\alpha_s$
       determinations, see, for example:
       B. Webber, ICHEP'94; I. Hinchliffe, DPF'94 and
       \PRD{50}{94}{1173}, p.1297.
\bibitem{us_mc} A. El-Khadra and B. Mertens, \NPBproc{42}{95}{406}.
\bibitem{elc_mq} C. Allton, {\em et al.}, \NPB{431}{94}{667}.

\bibitem{itep} V. Novikov, {\em et al.}, \PRepC{41}{78}{1}.
\bibitem{s-decays} R. Barbieri, G. Curci, E. d'Emilio and E. Remiddi,
       \NPB{154}{79}{535};
       K. Hagiwara, C. Kim, T. Yoshino, \NPB{177}{81}{461};
       P. Mackenzie and P. Lepage, \PRL{47}{81}{1244}.
\bibitem{p-decays} R. Barbieri, R. Gatto and R. K\"{o}gerler,
       \PLB{60}{76}{183};
       R. Barbieri, R. Gatto and E. Remiddi, \PLB{61}{76}{465};
       R. Barbieri, M. Caffo, R. Gatto and E. Remiddi, \PLB{95}{80}{93};
       \NPB{192}{81}{61}.
\bibitem{anl} G. Bodwin, S. Kim and D. Sinclair, \NPBproc{34}{94}{434};
       \NPBproc{42}{95}{306}.
\bibitem{mangano} M. Mangano and A. Petrelli, \PLB{352}{95}{445}.
\bibitem{ginsburg} C. Ginsburg (E760 collaboration), these proceedings.
\end{Thebibliography}

\end{document}